\documentclass[twoside,twocolumn,superscriptaddress,amsfonts,amssymb,lengthcheck,hidepacs,pra,aps,amsmath]{revtex4}

\usepackage{graphicx}
\usepackage{color}

\DeclareMathOperator{\erf}{erf}
\DeclareMathOperator{\tr}{Tr}
\begin{document}

\title{Theory of the Normal/Superfluid interface in population imbalanced Fermi gases}
\author{Stefan K. Baur}
\affiliation{Laboratory for Atomic and Solid State Physics, Cornell University, Ithaca, New York 14853,
USA.}
\author{Sourish Basu}
\affiliation{Laboratory for Atomic and Solid State Physics, Cornell University, Ithaca, New York 14853,
USA.}
\author{Theja N. De Silva}
\affiliation{Department of Physics, Applied Physics and Astronomy, SUNY Binghamton, Binghamton, NY 13902, USA.}
\author{Erich J. Mueller}
\affiliation{Laboratory for Atomic and Solid State Physics, Cornell University, Ithaca, New York 14853,
USA.}

\begin{abstract}
We present a series of theoretical studies of the boundary between a superfluid and normal region in a partially polarized gas of
strongly interacting fermions.  We present mean-field estimates of the surface energy in this boundary as a function of temperature and scattering length.  We discuss the structure of the domain wall, and  use a previously introduced phenomonological model to study its influence on experimental observables.
 Our microscopic mean-field calculations are not consistent with the magnitude of the surface tension found from our phenomonological modelling of data from the Rice experiments.  We conclude that 
 one must search for novel mechanisms to explain the experiments.
\end{abstract}
\maketitle
\section{Introduction}
What happens when one tries to polarize a fermionic superfluid?
Experiments at MIT and Rice have shown that when the fermions are interacting via resonant short range interactions, the fluid responds by
phase separating into a largely unpolarized superfluid region and a less polarized normal region \cite{Partridge2006,Partridge2006a,Zwierlein2006,Zwierlein2006a,Shin2008}.  The Rice experiments  \cite{Partridge2006,Partridge2006a} show a dramatic distortion of the central superfluid region in their trapped gas, pointing to significant surface tension in the boundary.  Here we present a theoretical study of this boundary and discuss the consequences of surface tension.

The phase separation seen in these experiments arises because a zero temperature conventional s-wave superfluid is unable to accommodate spin polarization:  all of the atoms in one spin state ($\uparrow$) are paired with atoms of the opposite spin ($\downarrow$).  Changing the density ratio $n_\uparrow/n_\downarrow$ from unity requires adding sufficient energy to break these pairs.  Consequently, when excess particles of one spin state are added to a paired atomic cloud, those particles simply float to the surface, forming a normal fluid.  Given that there is a sharp boundary between the superfluid and normal region, the order parameter must vary rapidly, producing a surface energy. In section~\ref{sec:micro} we use the Bogoliubov-de-Gennes(BdG) equations as a microscopic theory of this interface, and extract a dimensionless measure of the surface tension $\eta$.

In addition to the phase separation scenario seen in experiments, which was essentially predicted by Clogston and Chandrasekhar \cite{Clogston1962,Chandrasekhar1962}, there have been many theoretical proposals for how the superfluid can accomodate extra spins \cite{Fulde1964,Larkin1965,Gubankova2003,Deb2004,Gubankova2005,Shovkovy2003,Combescot2001,Lombardo2001,Caldas2004,Sedrakian2005,Son2006,He2006a,Yi2006a,Hu2006,Parish2007,Gubbels2006,Iskin2006,Yang2005,Chien2007,Parish2007a}.  We will make some comments on one class of states, the ``FFLO" states introduced by  Fulde, Ferrell, Larkin and Ovchinnikov \cite{Fulde1964,Larkin1965} where  polarization resides in a periodic array of nodes in the superfluid order parameter, or is accommodated by creating supercurrents.
Our mean field approach is sufficiently general that if such a phase existed in the parameter regime that we discuss, we could observe it.  Like the experiments, and previous mean-field calculations \cite{Yoshida2007}, we see no sign of this phase near unitarity (where the scattering length is infinite).  Interestingly, 
a recent density functional calculation \cite{Bulgac2008} raises the possibility that the mean-field theory may be underestimating the stability of the FFLO state.  We anticipate that in the near future more sophisticated numerical techniques will be able to unambiguously address the presence of the FFLO state.

%
Like other theoretical calculations based upon the Bogoliubov de Gennes equations\cite{Machida2006,Mizushima2005,Mizushima2007,Liu2007,Kinnunen2006a,Jensen2007,Sensarma2007,Castorina2005,Tezuka2008} we find that the order parameter and polarization oscillate in the domain wall separating the two regions.
These oscillations, which in no way should be interpreted as an intervening phase~\cite{Sensarma2007},  
are small at unitarity, but become larger as one approaches the BCS limit (small and negative scattering length $a$).
At sufficiently small negative scattering length the decay length of these oscillations diverge, signaling the onset of the bulk FFLO phase.
 The topology of the phase diagram of polarized fermions \cite{Parish2007,Sheehy2007} at zero temperature features tricritical points in both the BEC ($1/k_f a\gg1$) and BCS ($-1/k_f a\gg1$) limit, where the first order phase transition between superfluid and normal state turn second order. As surface tension vanishes at the tricritical points, it reaches a maximum in the crossover region.

In section~\ref{sec:theja1} we explore the consequences of surface tension on the shape of the superfluid-normal boundary for a unitary gas in an anisotropic harmonic trap.  As in previous work \cite{Silva2006a,Haque2007,Natu2008} we use a simple elastic model for the boundary.  By expanding the shape of the boundary in a Fourier series, we are able to compare its detailed structure with that of experiments.  While our model appears to capture a great many of the experimental features, it yields sharper density features.  We attribute the discrepancies to the fact that we model the trapping potential as harmonic.

This method complements the approach of Natu and Mueller \cite{Natu2008} where the conditions of hydrostatic equilibrium were used to derive a differential equation for the boundary.
Like Haque \textit{et al.} ~\cite{Haque2007}, we find that as one increases the surface tension, the boundary distorts from an ellipse into a ``capsule-like'' shape. The most relevant experiments,  performed at Rice in 2006 \cite{Partridge2006a}, use small numbers of particles in a highly anisotropic trap.  Due to the large surface area to volume ratio these experiments observe large distortions, consistent with surface tension which is one order of magnitude larger than predicted by our microscopic arguments. This behavior should be contrasted with experimental studies at MIT~\cite{Shin2008}  which find no observable distortion of the superfluid-normal boundary.  Taking into account the much smaller surface area to volume ratio in these experiments, this null observation bounds $\eta$ to be not much larger than the value we calculate.  We have no explanation for this apparent discrepency.  It is undoubtedly related to the fact that the Rice experiment finds a normal fluid whose local polarization is almost 100\%, while the normal fluid seen at MIT is always partially polarized, even at the lowest temperatures.

\subsection{Background}

In the experiments of interest, fermionic alkali atoms (typically $^6$Li) are trapped in a nominally harmonic optical potential.  The atoms are transfered into two collisionally stable hyperfine states so that no spin relaxation occurs on the timescale of the experiment:  the number of particles in each of the two spin states are separately conserved.  Due to the short range nature of the interactions, at sufficiently low temperature scattering is forbidden between two fermions in the same spin state.  Hence interactions can be parameterized by a single scattering length $a$, which is a function of the applied magnetic field \cite{Feshbach1958,Fano1961}.

For $a<0$ the low energy scattering is attractive, and at low temperature the spin balanced system is a BCS superfluid.  For $a>0$ the low energy scattering is repulsive, however there exists a two-body bound state in vacuum.  The low temperature phase in this case is a Bose condensate of pairs.  One of the remarkable predictions born out by experiments is that these two superfluid phases are continuously connected to one another \cite{Regal2004,Zwierlein2004,Chin2004,Bourdel2004,Kinast2004,Zwierlein2006a,Shin2008,huletmolecule}.
Most interest has focused around the unitary point ($a=\pm\infty$) where the scattering  cross-section is maximal and in free space there exists a bound state at threshold.  At this point the interactions do not provide a length-scale to the system, leading to universal thermodynamics where all thermodynamic functions can be expressed in terms of a power of the density times a universal function of the density scaled by the thermal wavelength $n\Lambda^3$ and the spin imbalance $n_\uparrow/n_\downarrow$ \cite{Heiselberg2001,O'Hara2002,Carlson2003,Bruun2004,Ho2004}.  In particular, if in the absence of a trap there is a flat phase boundary between a superfluid and a normal region, with equal pressures on each side of the boundary, we showed in~\cite{Silva2006a} that any surface tension can be written as
\begin{equation}
 \sigma = \eta\frac{\hbar^2}{2m}n_s^{4/3}
 \label{eq:defineeta}
\end{equation}
where $n_s$ is the density on the superfluid side of the boundary. A completely equivalent parameterization was later used by Haque and Stoof \cite{Haque2007}, 
\begin{equation}
\sigma=\eta_s \frac{m}{\hbar^2}\mu^2. 
\end{equation}
To convert between the two parameterizations one needs to know the equations of state of the superfluid -- which on dimensional grounds has the simple form 
\begin{equation}\label{sfeos}
n_s(\mu)=n_{\uparrow}+n_{\downarrow}=\frac{1}{3 \pi^2} [\frac{2m}{\hbar^2 (1+\beta)}\mu]^{3/2},
\end{equation}
 where $\mu$ is the average chemical potential and $\beta$ is a dimensionless universal many body parameter \cite{Heiselberg2001,O'Hara2002,Bruun2004,Ho2004}. According to quantum Monte-Carlo calculations $\beta \approx-0.58$ \cite{Carlson2003,Astrakharchik2004,Chang2004,Carlson2005,Lobo2006}, which gives $\eta=8.1\eta_s$.

If, as in the experiments, the boundary is curved, one expects to observe a pressure drop 
related to the curvature \cite{Natu2008}.  In that case, the dimensional argument leading to Eq.~(\ref{eq:defineeta}) yields an extra parameter, and one must take $\eta=\eta(\Delta p/p)$ to be  a function of the relative pressure drop.  As in all previous treatments, we will neglect the $\Delta p$ dependance of this parameter.

\begin{figure}[ht!]
	\centering
		\includegraphics[width=\columnwidth]{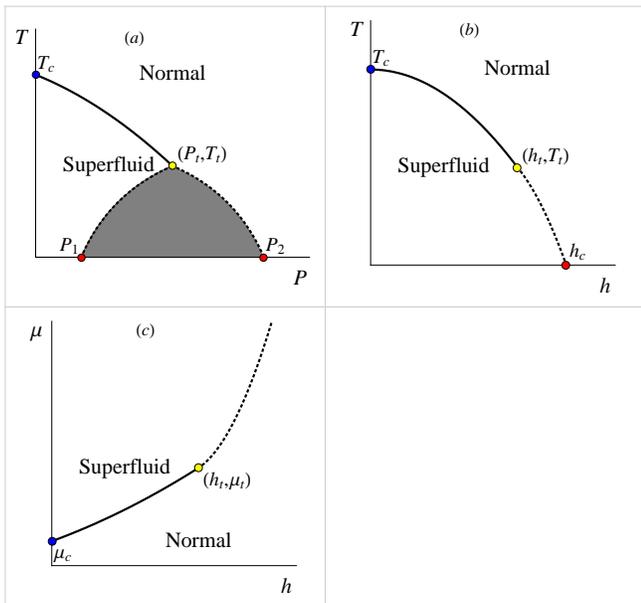}
	\caption{(Color Online) Schematic phase diagram of a two component Fermi gas as a function of (a) Temperature [$T$] - Polarization [$P=(n_\uparrow-n_\downarrow)/(n_\uparrow+n_\downarrow)$], (b)  Temperature [$T$] - chemical potential difference [$h=(\mu_\uparrow-\mu_\downarrow)/2$], and
	(c) chemical potential [$\mu=(\mu_\uparrow-\mu_\downarrow)/2$] -  chemical potential difference [$h=(\mu_\uparrow-\mu_\downarrow)/2$].   The equation of state sets a relationship between $P,T,h,$ and $\mu$, so only three of them are needed to specify the state.  Solid lines: continuous phase transitions; dashed lines: discontinuous; these meet at the tricritical point [$P_t,h_t,T_t$] or [$\mu_t,h_t,T_t$].  The gray region in (a) maps onto the dashed line in (b), and represents a coexistence region.
On the BCS side of resonance $a<0$, and for sufficiently large $a$, $P_1$=0 and $0<P_2<1$.  When $a>0$ decreases in magnitude, $P_1$ and $P_2$ move to the right, sequentially hitting the maximum allowed value $P=1$.   At unitarity, $a=\infty$, a Wilsonian RG theory\cite{Gubbels2008} predicts $P_t=0.24$ and $T_t/T_{F,\uparrow}=0.06$. Monte-Carlo calculations suggest $T_c/T_F=0.152(7)$\cite{Burovski2006}, and $P_2=0.39$\cite{Lobo2006}. }
	\label{fig:pdsketch1}
\end{figure}
The presence of a superfluid-normal phase boundary for the trapped gas is understood by examining
the finite temperature phase diagram of a uniform Fermi gas, as sketched in figure~\ref{fig:pdsketch1}(a) in the polarization-temperature plane. At zero temperature the superfluid and normal state are separated by a polarization driven first order phase transition.  Since the polarization changes discontinuously, there is a ``forbidden region" analogous to the one occuring in the density-temperature phase diagram for a liquid-gas phase transition.  The location of the phase boundaries are found by a standard Maxwell construction, where the pressures (free energies) of the two phases are equated. When placed in a fixed volume (or confined by a harmonic trap) this first order phase transition leads to a regime of phase coexistence, just as in the more familiar situation of a liquid-vapor transition.

This first order phase transition is only found at sufficiently low temperatures. There is a tricritical point, above which the boundary becomes a continuous second order line.  This behavior is consistent with the standard model of an unpolarized superfluid, which has a temperature driven second order phase transition. This structure was experimentally mapped out in \cite{Shin2008}.

In figure~\ref{fig:pdsketch1}(c) the same diagram is shown at fixed temperature as a function of the chemical potentials
$\mu=(\mu_\uparrow+\mu_\downarrow)/2$ and
 $h=(\mu_{\uparrow}-\mu_{\downarrow})/2$.  As a first approximation, one understands the trapped gas by breaking up the cloud into small pieces, and assuming that at each of these pieces is homogeneous and in local equilibrium.  Maintaining equilibrium in the presence of energy transport requires $T$ is constant, allowing momentum transport requires that the pressure obeys $\nabla P=-n \nabla V $, where $V(r)$ is the trapping potential, and allowing particle transport requires $\nabla \mu_j=-\nabla V $ for $j=\uparrow,\downarrow$.  This hydrodynamic, {\em Thomas-Fermi} description allows one to directly read the structure of the trapped gas from the homogeneous phase diagram in figure~\ref{fig:pdsketch1}(c): $h$ is constant and $\mu$ varies from a large value at the center of the cloud to a small value at the edge.  The iso-density contours of the cloud follow the iso-potential contours.

Observations at Rice are inconsistent with this local density approximation \cite{Silva2006}.   As already emphasized,
the missing element is that the surface energy described in Eq.~(\ref{eq:defineeta}) distorts the cloud.  Microscopically, this surface tension arises due to the energy cost of the superfluid-normal boundary, where the order parameter varies rapidly.  Depending on the size of the superfluid region it is energetically favorable to either shrink this boundary to reduce its area or, because the surface energy depends on density, shift the boundary to a low density region.  In  a spherical trap this effect changes the radius (and density) of the superfluid core.  In an elongated cloud, one generically expects that the aspect ratio of the superfluid region is reduced.
%
Phenomenological models based on this principle \cite{Silva2006a,Haque2007}, seem to account for the experimental observations.

By fitting various models to experimental density profiles, Haque and Stoof \cite{Haque2007} and DeSilva and Mueller \cite{Silva2006a} both made an estimate of $\eta$.  
The quoted values of $\eta$ in \cite{Silva2006a} were unintentionally scaled by a factor of $\hbar\omega_z/\mu_0$.  When corrected for this factor, they found that $\eta\approx0.6$ for the relatively high temperature data in \cite{Partridge2006}.  Haque and Stoof \cite{Haque2007} found $\eta\approx 4.8$ for the lower temperature data in \cite{Partridge2006a}.  This is consistent with the expectation that surface tension should drop as one approaches the tricritical point.  Here we compare our model with the data in \cite{Partridge2006a}, finding  $\eta\approx3$.  We attribute the slight discrepancy with Haque and Stoof to trap anharmonicities (which we did not include in our calculation).  We emphasize that $\eta\approx3$ is more than one order of magnitude larger than the value predicted by our microscopic model, $\eta\approx0.17$.

\section{Calculation of Surface Tension}
\label{sec:micro}
In this section we present a calculation of the surface tension in the BEC-BCS crossover. First we give a very crude order of magnitude estimate of the surface tension at $T=0$ in a unitary gas, then we use a numerical solution to the BdG equations to obtain a more rigorous result at zero temperature. In the deep BEC limit and at finite temperature one can use a simpler theory where the free energy is expanded in gradients of the order parameter. We use this theory to obtain the temperature dependece of the surface tension in the unitary limit and to test our numerical solution of the BdG equations in the BEC limit.

Related microscopic calculations have been performed by Caldas \cite{caldas2007} and Imambekov, Bolech, Lukin and Demler \cite{Imambekov2006}.

\subsection{Order of magnitude}\label{sec:stefan1}
Before presenting a detailed microscopic calculation of the surface tension we give a simple estimate of its magnitude at zero temperature in a unitary gas.  In the standard semi-phenomonological model for surface tension, one considers the spatial variation of an order parameter: in this case the superconducting gap $\Delta({\bf r})$.
At the first order phase boundary, the free energy has two local minima: one with $\Delta=\Delta_0$ and the other with $\Delta=0$.  Surface tension can be attributed to the fact that in the boundary between the two bulk phases, $\Delta$ must pass over a free energy maximum.

The energy cost per unit area of the boundary $\sigma$ is most simply estimated as the product of the maximum height of the free energy barrier per unit volume $\delta \Omega$ and the thickness of the domain wall $\xi$.  The healing length $\xi$ arises from a competition between the stiffness of the order parameter and the height of the energy barrier, and at unitarity should be of order the interparticle spacing. Within BCS theory (reviewed in detail below), the energy barrier is $\delta \Omega\approx 0.2 \times n_s^{5/3} \hbar^2/2m$. With $\xi\approx n_s^{-1/3}$ this gives $\eta\approx 0.2$. As previously quoted, the full solution of the Bogoliubov de Gennes equations yield $\eta\approx 0.17$.

\begin{figure*}
	\centering
	\includegraphics[width=\textwidth,clip]{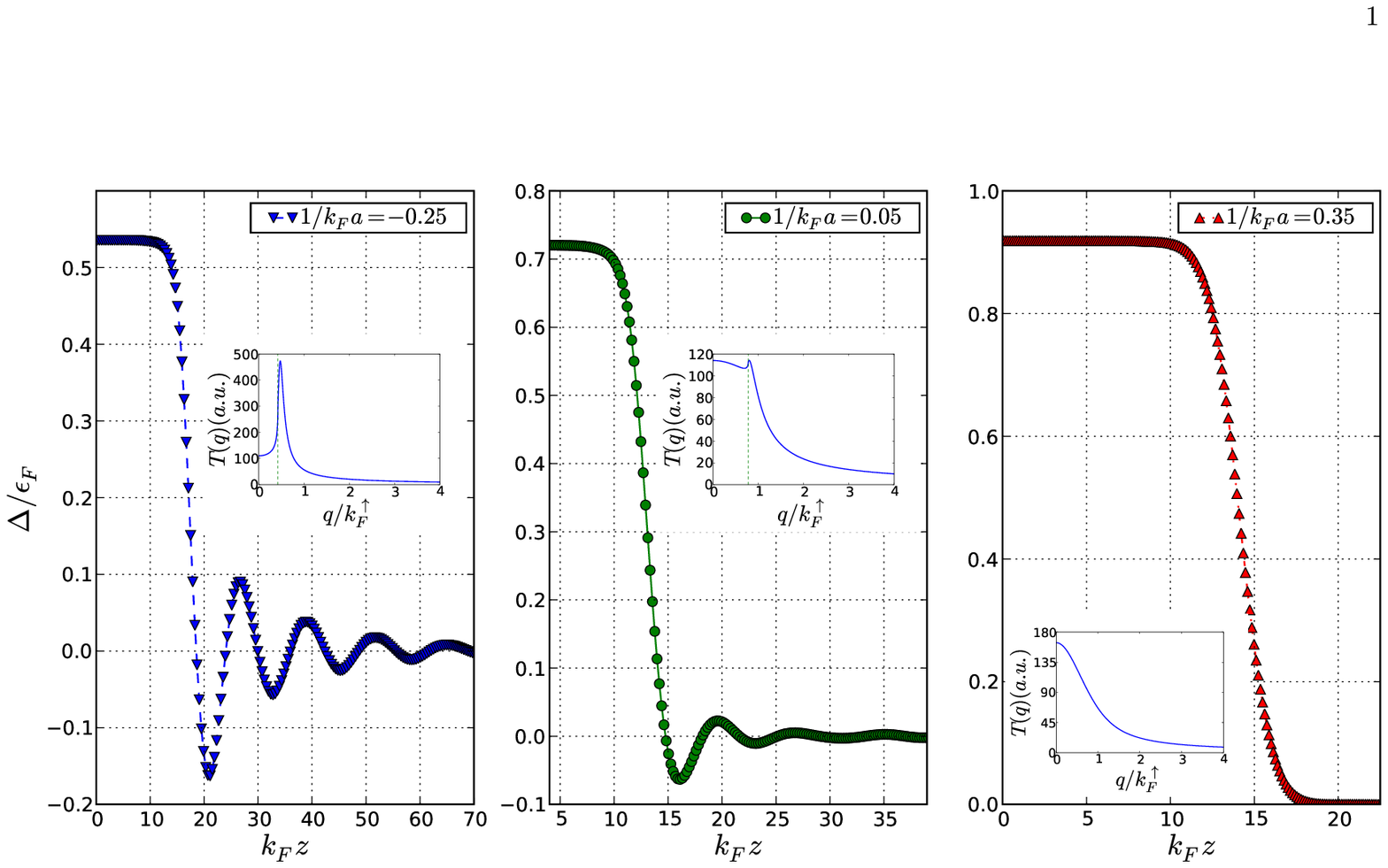}
	\caption{(Color Online) Order parameter profiles at the interface between normal and superfluid at critical Zeeman field $h_c$.  Left to right:  BCS to BEC side of resonance. Each data point corresponds to a single gridpoint of our real space discretization. Insets: normal state $T$-matrix (pair susceptability) as a function of momentum $q$  at the first order phase transition line $h=h_c$ corresponding to the same parameters as the BdG calculations. The Fourier transform of $T(q)$ describes the decay of the superfluid order parameter into the polarized normal state. The vertical line shows $q=k_F^{\uparrow}-k_F^{\downarrow}$.}
	\label{fig:dwplot}
\end{figure*}

\subsection{Mean Field Theory}\label{sec:sourish1}
To calculate the properties of the superfluid-normal boundary, we numerically solve the Bogoliubov de Gennes  equations with appropriate boundary conditions.
 The homogeneous version of these equations is often used to describe the zero temperature BEC-BCS crossover in ultracold fermions. We emphasize however that near unitarity, quantitative predictions of this theory should be viewed with some skepticism.  
  One could argue that since the mean-field theory overestimates the density discontinuity at the superfluid-normal boundary it should also overestimate $\eta$.
 In this section we review the formal theory, while in the following sections we report results, and describe some simplifying approximations.

Our model consists of a Fermionic system with two different hyperfine states
labeled by $\sigma = \uparrow, \downarrow$ in three spatial
dimensions. The atoms interact via a point interaction.
The
system of $N=N_{\uparrow}+N_{\downarrow}$ atoms is then described by the
Hamiltonian $H=\int d^3{\bf r} (H_0+H_{\rm int})$, with kinetic ($H_0$) and interaction ($H_{\rm int}$) energy densities
\begin{equation}
\begin{split}
H_0(\mathbf{r}) &= \sum_{\sigma} \psi^{\dagger}_{\sigma}(\mathbf{r}) \big(-\frac{\hbar^2}{2 m}\nabla^2-\mu_{\sigma} \big) \psi_{\sigma}(\mathbf{r})\\
H_{\rm int}(\mathbf{r}) &= -U \psi^{\dagger}_{\uparrow}(\mathbf{r})\psi^{\dagger}_{\downarrow}(\mathbf{r})\psi_{\downarrow}(\mathbf{r})\psi_{\uparrow}(\mathbf{r})
\end{split}
\end{equation}
where $\psi_{\sigma}(\mathbf{r})$ are usual Fermionic field
operators. Parameters $m$, and $\mu_{\sigma}$ are the mass and
chemical potential of the atomic species $\sigma$ respectively. Following convention, we
take $\uparrow$ to be the majority species of atoms and use variables
$h=(\mu_{\uparrow}-\mu_{\downarrow})/2\geq 0$ and 
$\mu=(\mu_{\uparrow}+\mu_{\downarrow})/2$. The
bare coupling constant $U$ is renormalized by expressing it through
the s-wave scattering length $a_s$ as $1/U=1/U_R+1/V
\sum_{q} 1/2 \epsilon^0_q$ with $U_R=-4 \pi \hbar^2 a_s/m$,
$\epsilon^0_q=\hbar^2 q^2/2 m$ where $V$ is the system
volume\cite{Ohashi2003}.

Performing a mean-field decoupling of the interaction, one writes the Hamiltonian in terms of a gas of Bogoliubov excitations \cite{Gennes2001},
\begin{equation}
\begin{pmatrix}
\Psi_{\uparrow}(\mathbf{r})
\\
\Psi_{\downarrow}^{\dagger}(\mathbf{r})
\end{pmatrix}
=\sum_n
\begin{pmatrix}
u_{n}(\mathbf{r}) & -v^{*}_{n}(\mathbf{r}) \\
v_{n}(\mathbf{r}) & u^{*}_{n}(\mathbf{r})
\end{pmatrix}
\begin{pmatrix}
\gamma_{\uparrow,n}
\\
\gamma^{\dagger}_{\downarrow,n}
\end{pmatrix}
\end{equation}
Excitations with spin $\sigma=\uparrow,\downarrow$ have energies $E_{n,\sigma}=E_n\pm h$, where $E_n$ is the positive energy eigenvalue of
the BdG equations
\begin{equation}
\label{BdGequations}
\begin{pmatrix}
-\frac{\nabla^2}{2 m}-\mu& \Delta(\mathbf{r}) \\
\Delta^*(\mathbf{r}) & \frac{\nabla^2}{2 m} +\mu
\end{pmatrix}
\begin{pmatrix}
u_{n}(\mathbf{r}) \\
v_{n}(\mathbf{r})
\end{pmatrix}
=E_{n}
\begin{pmatrix}
u_{n}(\mathbf{r}) \\
v_{n}(\mathbf{r})
\end{pmatrix}
\end{equation}
The mean-field free energy is then
\cite{Bardeen1969,Eilenberger1965}
\begin{eqnarray}
\label{mean_field_free_energy}
\Omega_{MF}&=&\sum_{n}\left\{\epsilon_{n}-\mu-\lambda_n\right\} 
+\int d^3\mathbf{r} \frac{ |\Delta(\mathbf{r})|^2}{U}\\\nonumber
\frac{\lambda_n}{k_B T}&=&\log\left(2\cosh  \frac{\beta E_{n,\uparrow}}{2} \right)+\log\left(2 \cosh  \frac{\beta E_{n,\downarrow}}{2}\right)
\end{eqnarray}
where $\sum_{n}\epsilon_n-\mu=\tr\left[-\frac{\nabla^2}{2 m}-\mu \right]$ and $\beta=1/k_B T$.
The chemical potentials are set by the number equations $N_{\sigma}=\int d^3r\,n_{\sigma}(\mathbf{r})$ with
\begin{eqnarray}
 n_{\sigma}(\mathbf{r}) &=& \left\langle  \psi^{\dagger}_{\sigma}(\mathbf{r})\psi_{\sigma}(\mathbf{r})\right\rangle\\\nonumber
 &=& \sum_n |u_n(\mathbf{r})|^2 f(E_{n,\sigma})+|v_n(\mathbf{r})|^2\left[ 1-f(E_{n,-\sigma})\right]
\end{eqnarray}
where $f(E)=1/(1+e^{-\beta E})$. Self-consistency requires that the gap obeys
\begin{equation}
\label{gapeq}
\begin{split}
\Delta(\mathbf{r}) &= U \left\langle \psi_{\uparrow}(\mathbf{r})\psi_{\downarrow}(\mathbf{r}) \right\rangle\\
&=U \sum_n u_n(\mathbf{r})v^{*}_n(\mathbf{r})\left[ 1-f(E_{n,\uparrow})- f(E_{n,\downarrow})\right].
\end{split}
\end{equation}
It is useful to draw attention to  three particular limits of these equations.  First, if $\Delta({\bf r})=0$ one has just a non-interacting Fermi gas, with free energy
\begin{equation}
\Omega=-\sum_{\mathbf{k},\sigma} k_B T \log \left( 1+e^{-\beta (\epsilon_k-\mu_{\sigma})}\right).
\end{equation}
This equation highlights the most significant flaw of the mean-field approach, namely that it yields a noninteracting normal state,  overestimating the stability of the superfluid.
 
As a second useful limit, one can consider the case where $\Delta({\bf r})=\Delta_0$ is uniform.  In that case one can label the eigenstates of (\ref{BdGequations}) by momentum and one finds the standard result
\begin{eqnarray}
\label{homogenous_mean_field_free_energy}
\Omega_{\rm hom}&=&\sum_{\mathbf{k}}  \left\{ \epsilon_{\mathbf{k}}-\mu-\lambda_k \right\} + V \frac{\Delta_0^2}{U}\\\nonumber
\frac{\lambda_k}{k_B T}&=& \log \left(2 \cosh \frac{\beta (E_{\mathbf{k}}+h)}{2}\right) \\\nonumber
&&\quad +\log \left(2\cosh \frac{\beta (E_{\mathbf{k}}-h)}{2}\right) 
\end{eqnarray}
where $V$ is the system volume and $E_{\mathbf{k}}=\sqrt{(k^2/2m-\mu)^2+\Delta_0^2}$. In this uniform limit one finds that 
the total
particle number $N=N_{\uparrow}+N_{\downarrow}$, population
imbalance $\Delta N=N_{\uparrow}-N_{\downarrow}$ and gap equation
become
\begin{widetext}
\begin{eqnarray}
N&=&-\frac{\partial \Omega}{\partial \mu}=\sum_{\mathbf{k}}\left(1-\frac{\epsilon_\mathbf{k}-\mu}{2 \sqrt{(\epsilon_{\mathbf{k}}-\mu)^2+\Delta_0^2}} \left\{
\tanh{\frac{\beta(E_{\mathbf{k}}-h)}{2}}+\tanh{\frac{\beta(E_{\mathbf{k}}+h)}{2}} \right\} \right) \label{eq:totalspins}\\
\Delta N&=&-\frac{\partial \Omega}{\partial h}=\frac{1}{2} \sum_{\mathbf{k}} \left(\tanh\frac{\beta(E_{\mathbf{k}}+h)}{2}-\tanh\frac{\beta(E_{\mathbf{k}}-h)}{2}\right)
\\
0&=&\frac{\partial \Omega}{\partial \Delta_0}=-\sum_{\mathbf{k}}\Delta_0 \frac{\frac{1}{2}\left( \tanh\frac{\beta(E_{\mathbf{k}}+h)}{2} +\tanh\frac{\beta(E_{\mathbf{k}}-h)}{2}\right)}{\sqrt{(\epsilon_{\mathbf{k}}-\mu)^2+\Delta_0^2}}+2 V \frac{\Delta_0}{U}\label{eq:gap_equation}
\end{eqnarray}
\end{widetext}

A third important case is when $\Delta({\bf r})$ is periodic.  Prototypical examples are the FF state $\Delta=\Delta_{FF} e^{i q x}$ or the LO state $\Delta=\Delta_{LO}\cos(q x)$.  In the BCS limit there is a small window of stability for such a solution. 
These exotic superfluid states can be described by the BdG approach.

In order to compute the surface tension across the BEC-BCS crossover we  numerically solve
the   BdG equations in  Eq.~(\ref{BdGequations}). We find the parameters such that bulk normal and superfluid have the same free energy and then minimize the free energy functional in Eq. ~(\ref{mean_field_free_energy}) with respect to the order parameter for a domain wall between the two phases. A simple way to minimize the free energy functional is to iterate the gap equation Eq.~(\ref{gapeq}) to self consistency.
We find that it is more computationally efficient to directly minimize
Eq.~(\ref{mean_field_free_energy}) with respect to a discretized representation of  $\Delta(r)$.  For efficiency we calculate the gradient using  
\begin{eqnarray}
\frac{\delta \Omega_{MF}[\Delta(\mathbf{r})]}{\delta \Delta(\mathbf{r})} &=& -\frac{2}{U} \left[\Delta(\mathbf{r})-\tilde\Delta(\mathbf{r})\right]\\\nonumber
\frac{1}{U}\tilde\Delta(\mathbf{r})&=& \sum_n u_n(\mathbf{r})v^{*}_n(\mathbf{r})\left[ 1-f(E_{n,\uparrow})- f(E_{n,\downarrow})\right]
\end{eqnarray}
where we used $\delta E_n/\delta \Delta(\mathbf{r})=2 u_n(\mathbf{r}) v_n(\mathbf{r})$.

\subsection{Results ($T=0$)}\label{sec:bdg}
The  domain wall profiles calculated within the BdG approach  are shown in figure~\ref{fig:dwplot}.  As can be seen, on the BEC side of resonance the domain wall is largely featureless, while oscillatory structures develop as one approaches the BCS limit.  In section~\ref{proximity} we relate these features to the behavior of the $T$-matrix, and draw the connection between these oscillations and the FFLO phase.

\begin{figure}[ht!]
	\centering
		\includegraphics[width=\columnwidth]{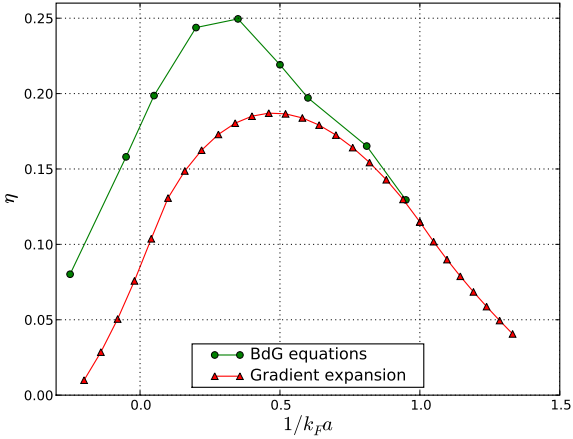}
	\caption{(Color Online) Dimensionless surface tension $\eta=2 \hbar^{-2} m n_s^{-4/3}  \sigma$ as a function of $(k_F a)^{-1}$ at $T=0$. When $(k_F a)^{-1}>1.01$ the superfluid state is partially polarized. Triangles: calculation using the full BdG equations as described in~\ref{sec:bdg}, circles: gradient expansion approximation to this solution from ~\ref{sec:sourish2}. The lines are a guide to the eye.}
	\label{fig:st}
\end{figure}

Due to the contribution from the interface, the energy found in these calculations exceeds that of the bulk superfluid/normal gas.   From this excess energy 
 we extract the dimensionless parameter $\eta$; our results are summarized in Fig.~\ref{fig:st}.
 At unitarity we find $\eta \simeq 0.17$.  The surface tension drops as one approaches the BCS side of resonance.  It grows to a maximum of $\eta \simeq 0.25$ at $1/k_f a\approx 0.4$, then falls as one proceeds towards the BEC limit.

\subsection{Gradient expansion}\label{sec:sourish2}
Even in the variational form discussed in the previous section, solving the BdG equations is numerically challenging -- especially at finite temperature.  To explore how the surface tension varies with temperature we consider a Landau-Ginzburg expansion,
\begin{equation}
\begin{split}
\Omega =& \int d^3 r\,\Bigl[\alpha_1\left|\Delta(r)\right|^2 + \alpha_2\left|\nabla\Delta(r)\right|^2 + \alpha_3\left|\nabla^2\Delta(r)\right|^2 + \cdots\Bigr.\\
&+ \gamma_1\left|\Delta(r)\right|^4 + \gamma_2\left|\Delta(r)\right|^6 + \cdots\\
&+ \Bigl.\text{gradient terms of order higher than } \Delta^2 \Bigr]
\end{split}
\label{eq:LGfull}
\end{equation}
The sum of the terms $\alpha_1$, and  $\gamma_1$, $\gamma_2, \cdots \gamma_n$ yields the homogeneous free energy in Eq.~\eqref{homogenous_mean_field_free_energy}, and we include all of them by numerically calculating $\Omega_{\rm hom}$. Additionally, in this section we calculate the gradient terms which are quadratic in $\Delta$, namely $\alpha_2$, $\alpha_3, \cdots$.  We will neglect gradient terms higher than $\mathcal{O}(\Delta^2)$, an approximation well-justified near the tricritical point.  By comparing to our solutions of the BdG equations, we find that this approximation introduces about a factor of 2 error at zero temperature. 
For the terms which are quadratic in $\Delta$, we go to all orders in the gradient.  At low $T$ the first term, $\alpha_2$, becomes negative and higher order terms stabilize the system.

This expansion is readily developed from the path integral representation of the partition function, after decoupling the interaction term with a Hubbard-Stratonovich transformation
\begin{widetext}
\begin{equation}
\begin{split}
S =& \underbrace{\int_0^\infty d\tau\sum_{k\sigma}\psi^*_{k\sigma}\left(\partial_\tau + \epsilon_k-\mu_\sigma\right)\psi_{k\sigma}}_{S_0} + \frac{1}{U}\int_0^\beta d\tau\sum_q\left|\Delta_q\right|^2\\
&+ \underbrace{\int_0^\beta d\tau\sum_q\Delta_q^*\sum_k\psi_{\frac{q}{2}-k\downarrow}\psi_{\frac{q}{2}+k\uparrow} + \int_0^\beta d\tau\sum_q\Delta_q\sum_k\psi^*_{\frac{q}{2}+k\uparrow}\psi^*_{\frac{q}{2}-k\downarrow}}_{S_1}\\
Z \propto& \int\left\{\prod_p d\Delta_p\,d\Delta^*_p\right\} e^{-\frac{1}{U}\int_0^\beta d\tau\sum_p\left|\Delta_p\right|^2} \times \underbrace{\int\left\{\prod_{k\sigma}d\psi_{k\sigma}d\psi^*_{k\sigma}\right\}e^{-S_0}e^{-S_1}}_{\left\langle e^{-S_1}\right\rangle} \\
\end{split}
\end{equation}
\end{widetext}
where the expectation $\langle e^{-S_1}\rangle$ is evaluated
in the free fermion ensemble. Evaluating this expectation yields the
partition function in terms of the pair susceptibility $\chi(q)$:
\begin{equation}
\begin{split}
Z &\propto \int\left\{\prod_p d\Delta_p\,d\Delta^*_p\right\} e^{\beta\sum_q\chi(q)\left|\Delta_q\right|^2 + \mathcal{O}[\Delta^4]} \\
\chi(q) &= \sum_k \left\{ \frac{1-f(\epsilon_{\frac{q}{2}+k\uparrow})-f(\epsilon_{\frac{q}{2}-k\downarrow})}{\epsilon_{\frac{q}{2}+k\uparrow}+\epsilon_{\frac{q}{2}-k\downarrow}} - \frac{1}{2\epsilon_k} \right\}
\end{split}
\label{eq:definechi}
\end{equation}
The function $\chi(q)$ is of general importance in the many-body problem of the BEC-BCS crossover of spin polarized fermions, such as T-Matrix approximation schemes~\cite{Ohashi2003} and the finite-momentum pairing instability of the polarized normal phase to the FFLO phase~\cite{Casalbuoni2004,Parish2007}. It is the static limit of the more general two-particle propagator $\chi(\mathbf{q},i\nu_n)$
\begin{equation}
 \chi(\mathbf{q},i\nu_n) = \sum_\mathbf{k}\left\{\frac{1 - f\left(\epsilon_{\frac{\mathbf{q}}{2}+\mathbf{k},\uparrow}\right) - f\left(\epsilon_{\frac{\mathbf{q}}{2}-\mathbf{k},\downarrow}\right)}{\epsilon_{\frac{\mathbf{q}}{2}+\mathbf{k},\uparrow} + \epsilon_{\frac{\mathbf{q}}{2}-\mathbf{k},\downarrow} - i\nu_n} - \frac{1}{2\epsilon_\mathbf{k}}\right\}
\end{equation}
At zero temperature the integrals evaluate to
\begin{eqnarray}
2\pi \tilde{\chi}(\mathbf{q},\nu) &=& 2\pi i\zeta - 2(1+r)-\frac{f_+^\uparrow+f_-^\uparrow+f_+^\downarrow+f_-^\downarrow}{\tilde q}\\\nonumber
f_\sigma^\uparrow &=&
\left(\frac{{\tilde q}^2}{4}+\zeta^2-1+\sigma \tilde q \zeta\right) \log\left(
\frac{\tilde q -2 +2\sigma \zeta}{\tilde q +2 +2\sigma\zeta}\right)\\\nonumber
f_\sigma^\downarrow &=&
\left(\frac{{\tilde q}^2}{4}+\zeta^2-r+\sigma \tilde q \zeta\right) \log\left(
\frac{\tilde q -2 r+2\sigma \zeta}{\tilde q +2r +2\sigma\zeta}\right)
\end{eqnarray}
where $\tilde{\chi}=(4\pi/mVk_F^\uparrow)\chi$, $\tilde{q}=\left|\mathbf{q}\right|/k_F^\uparrow$, $r=k_F^\downarrow/k_F^\uparrow$, $\tilde{\nu}=m\nu/(k_F^\uparrow)^2,\zeta = \tilde{\nu}+(1+r^2)/2-\tilde{q}^2/4$ and $k_F^{\uparrow,\downarrow}=\sqrt{2 m (\mu \pm h)}$, and for brevity we have set $\hbar=1$. When $\mu_{\uparrow,\downarrow}\leq0$, one takes $k_F^{\uparrow,\downarrow}\equiv0$.

The normal state becomes locally unstable towards finite $q$ pairing when the coefficient of $|\Delta_q|^2$ in  Eq.~(\ref{eq:definechi}) becomes negative:
\begin{equation}
 \chi(\mathbf{q},0) - \frac{1}{U} = 0.
\end{equation}
Above the tricritical point this
Thouless criterion locates the second order phase boundary, while below the tricritical point it yields the spinodal of the first order phase transition.

\subsubsection{Mean Field approximation}
At this point we neglect fluctuations in $\Delta$, finding that to quadratic order
\begin{equation}
Z=e^{-\beta\Omega} \approx \exp\left(-\beta\sum_q T^{-1}(q) |\Delta_q|^2\right),
\end{equation}
where the $T$-matrix obeys $T^{-1}(q)=1/U-\chi(q)$.  Thus explicit expressions for
 $\alpha_1$, $\alpha_2$, $\alpha_3, \cdots$ are found by expanding $\chi(q)$ in powers of $q$.
 
 Using this expression for the quadratic terms in (\ref{eq:LGfull}), we get
 \begin{equation}\label{eq:om1}
 \Omega[\Delta]\approx\int d^3r\,\frac{\Omega_{\rm hom}(\Delta(r))}{V}+\sum_q (\chi(0)-\chi(q))|\Delta_q|^2,
 \end{equation}
 where $\Delta(r)=V^{-1}\sum_q e^{i{\bf q\cdot r}} \Delta_q$.  The $q=0$ term is subtracted to eliminate double-counting of the $\alpha_1$ term.
 
 We calculate the surface energy from~\eqref{eq:om1}  by employing the ansatz
 \begin{equation}
\Delta(z)=\Delta_0 (\erf(4 z/W_{\rm dw})+1)/2
\label{eq:erf_ansatz}
\end{equation}
and minimizing $\Omega$ with respect to $W_{\rm dw}$, which is a measure of the width of the domain wall.  Technical details are describe in appendix~\ref{sourish_appendix}.  Figure~\ref{fig:fig2} shows the temperature dependance of the domain wall width at unitarity.  As can be seen, the domain wall size diverges at the tricritical point.
\begin{figure}[ht!]
\centering
\includegraphics[width=\columnwidth]{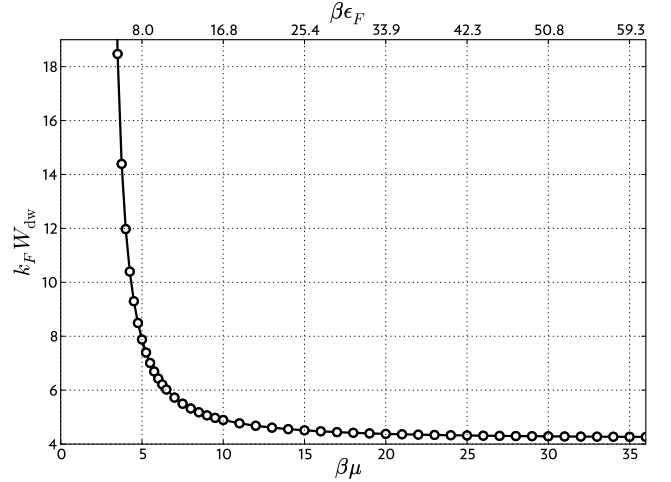}
\caption{Width ($W_{dw}$) of the domain wall as function of inverse
temperature ($\beta$) at unitarity, with parameters measured in terms of the chemical potential $\mu$ and $k_F^3 = 3\pi^2 n_s$.  Top scale is nonlinear.
The 
domain wall width diverges at the 
tricritical point  around $\beta\mu\sim
2.0$.}
\label{fig:fig2}
\end{figure}

\begin{figure}[ht!]
\centering
\includegraphics[width=\columnwidth]{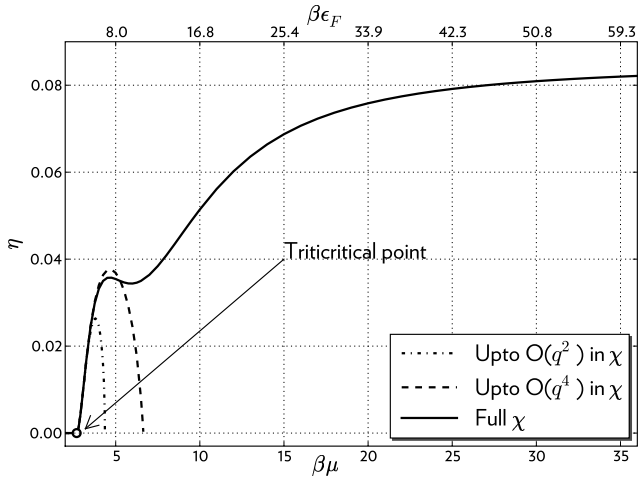}
\caption{Dimensionless surface tension $\eta=2 \hbar^{-2} m n_s^{-4/3}  \sigma$ as a function of inverse
temperature $\beta$, keeping different numbers of terms in the gradient expansion.    Even with the full $\chi$, we include terms up
to ${\cal O}(\Delta^2)$ in the free energy. As temperatures goes to
zero, the solid line suggests $\eta\sim 0.1$.  This should be compared with the full solution of the Bogoliubov-de-Gennes equations which give $\eta=0.17$.
The kink near $\beta\mu\approx6$ is an artifact of neglecting gradient terms which are beyond quadratic order in $\Delta$.
} \label{fig:fig1}
\end{figure}

The surface tension $\eta$, extracted from the free energy evaluated at the optimal $W_{\rm dw}$, is shown in figure~\ref{fig:fig1}.  We also show what this surface tension would be if we
expand $\chi(q)$ to either quadratic or quartic order in $q$.  This corresponds to truncating (\ref{eq:LGfull}) at $\alpha_2$ or $\alpha_3$. While these latter approximations work well around the tricritical point, they do not correctly describe the low temperature physics:  both $\alpha_2$ and $\alpha_3$ change sign at low temperature, and without the influence of higher order terms, the normal state becomes unstable to a FFLO state, and the surface tension vanishes.  Using the full $\chi$, we find that as $T\to0$, the dimensionless surface tension becomes $\eta\sim0.1$.  The discrepency between this result and the full BdG equations can be attributed to the neglect of gradient terms which are higher order in $\Delta$.  

Comparing these gradient expansion results to the full solution to the BdG equations at  $T=0$, we find that the agreement becomes better in the deep BEC limit
(see Fig.~\ref{fig:st}). 
In this limit, $\chi$ is well approximated by a parabola, and $\Omega_{\rm hom}$ can be truncated at quartic order in $\Delta$.  Minimizing the resulting free energy results in a  Gross-Pitaveskii equation~\cite{Pieri2006}.  Recently, Sheehy \cite{Sheehy2008} has discussed the role of quantum fluctuations near the tricritical point.

\subsubsection{Proximity effect}\label{proximity}
On the normal side of the domain wall, where $\Delta(z)$ is small, the free energy can be expanded to quadratic order in $\Delta$.  To find the asymptotic behavior of $\Delta$ in the domain wall one minimizes the quadratic approximation, $\Omega\approx \sum_{\bf q} T^{-1}({\bf q}) |\Delta_{\bf q}|^2$ with the constraint that $\Delta(z=0)=\bar \Delta$ is nonzero.  By symmetry, $\Delta_{\bf q}=\Delta_{q_z}$, and one finds
\begin{equation}
T^{-1}(q_z) \Delta_{q_z}-\lambda=0,
\end{equation}
where $\lambda$ is a Lagrange multiplier.  Consequently, for large $z$, the order parameter is proportional to the {\em one dimensional} Fourier transform of the T-matrix
\begin{equation}
\Delta(z) \propto \int \frac{dq}{2 \pi}  e^{i q z} T(q).
\end{equation}
This result is confirmed in figure~\ref{fig:dwplot}, where insets show the behavior of $T(q)$.  For example, on the BEC side of resonance the T-matrix is peaked at $q=0$, yielding a monotonically decaying order parameter.  On the BCS side, the T-matrix is peaked at finite $q\approx k_F^\uparrow-k_F^\downarrow$, and one observes oscillations of $\Delta(z)$ with this wave-vector.  

One can make a simple analogy, noting that the q-dependance of the T-matrix is reminiscent of the frequency dependance of  a driven damped harmonic oscillator.  The spatial dependance of the order parameter domain wall would then be analogous to the temporal dependance of the oscillator's position once the drive is turned off.  The BEC/BCS sides of resonance then map on to overdamped/underdamped oscillators.


\section{Effect of surface tension on density profiles}\label{sec:theja1}

\begin{figure*}[htbp]
 \includegraphics[width=\textwidth]{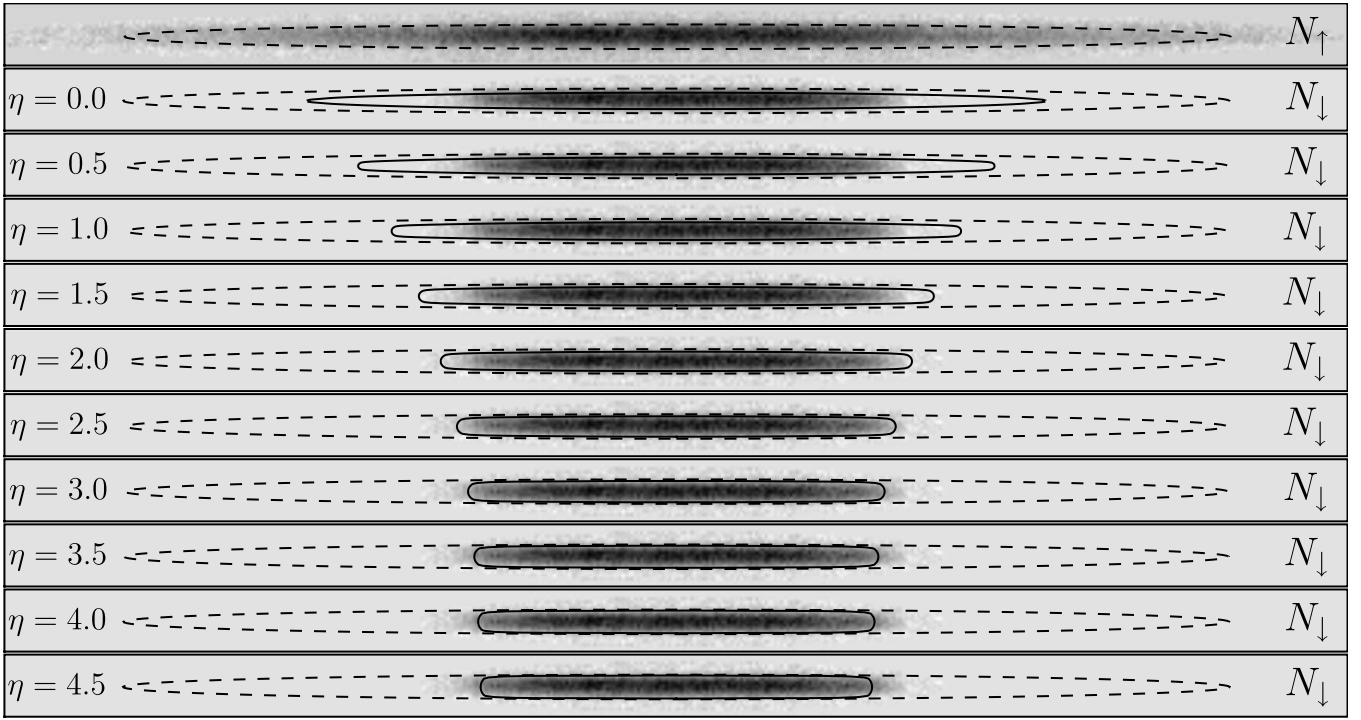}
 \caption{Experimental two-dimensional column densities (black denotes high density) for $P=0.38$ with theoretically calculated boundaries for different surface tensions $\eta$ (fixing the number of particles to be constant). 
 Top: majority atoms $N_\uparrow$; Bottom: minority atoms $N_\downarrow$.
 The dotted line is the ellipse with semi-major and semi-minor axes $Z_{\rm TF}$ and $R_{\rm TF}$ respectively, while the solid line is the superfluid-normal boundary in the presence of surface tension. As $\eta$ is increased, the superfluid-normal boundary deforms from an elliptical isopotential surface, but the boundary becomes increasingly insensitive to surface tension with increasing $\eta$. $N_c=15$ Fourier components were chosen for equation~\eqref{eq:harmonicansatz}. Data corresponds to Fig.~1(c) in Ref.~\cite{Partridge2006a}, used with permission.  Data outside of an elliptical aperture has been excluded.  This truncation of the data leads to a slight discrepancy in $P$ compared to the value quoted in \cite{Partridge2006a}. Each panel is $1.4$mm$\times0.06$mm, and shows the true aspect ratio of the cloud.
 }
 \label{fig:inceta}
\end{figure*}

\begin{figure*}[htbp]
 \includegraphics[width=0.48\textwidth]{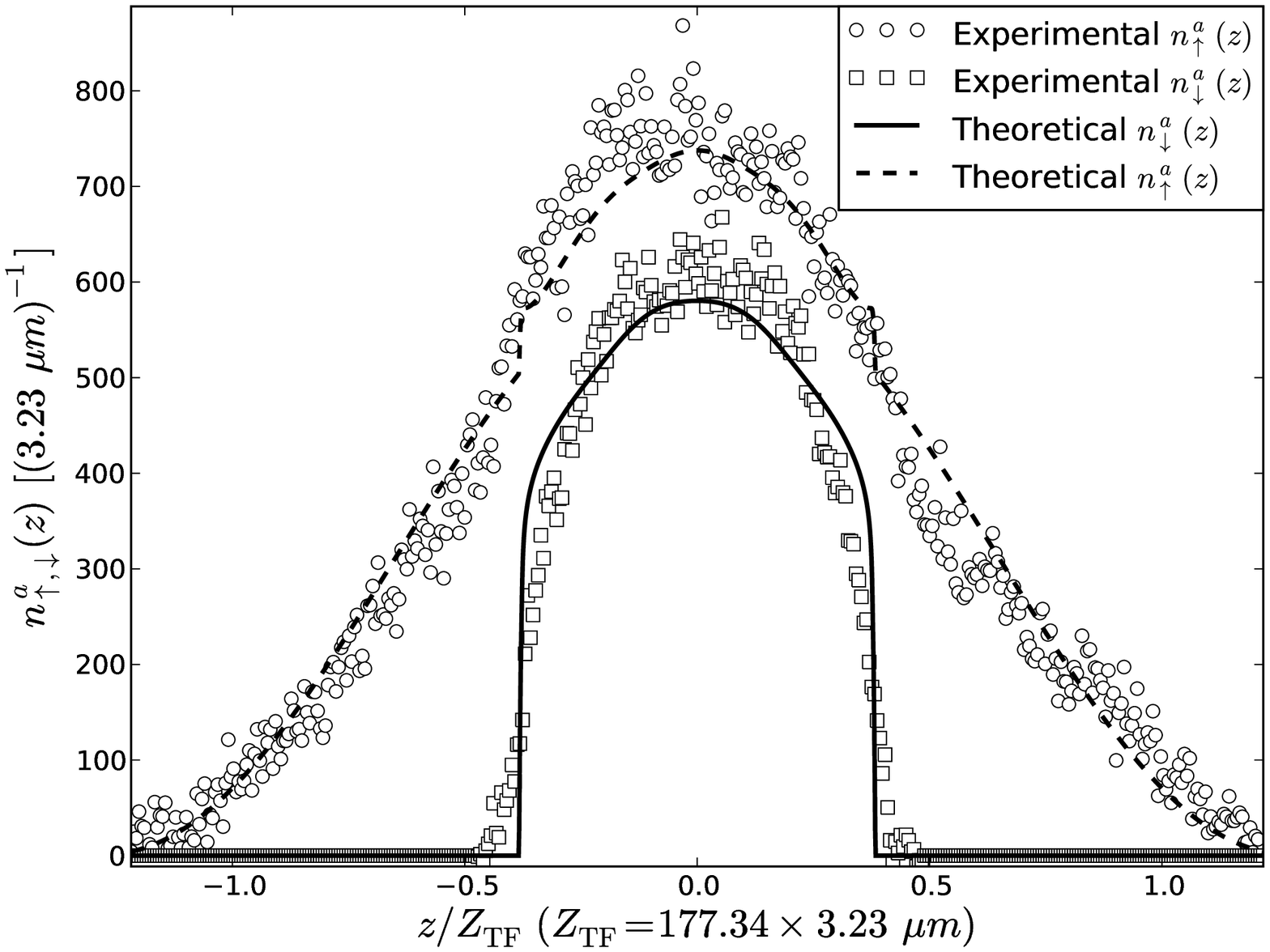}
 \hfill
 \includegraphics[width=0.48\textwidth]{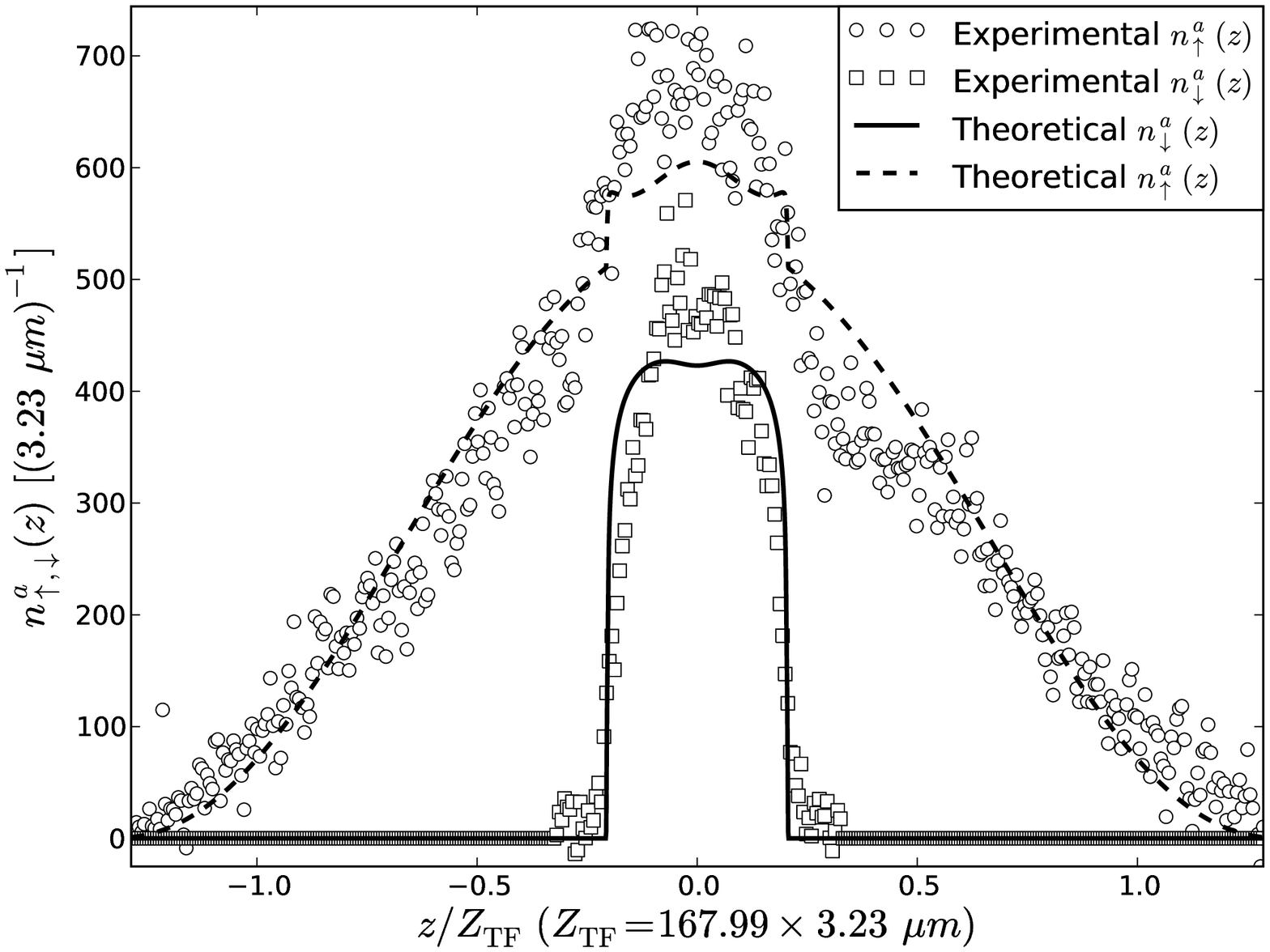}\\
 \includegraphics[width=0.48\textwidth]{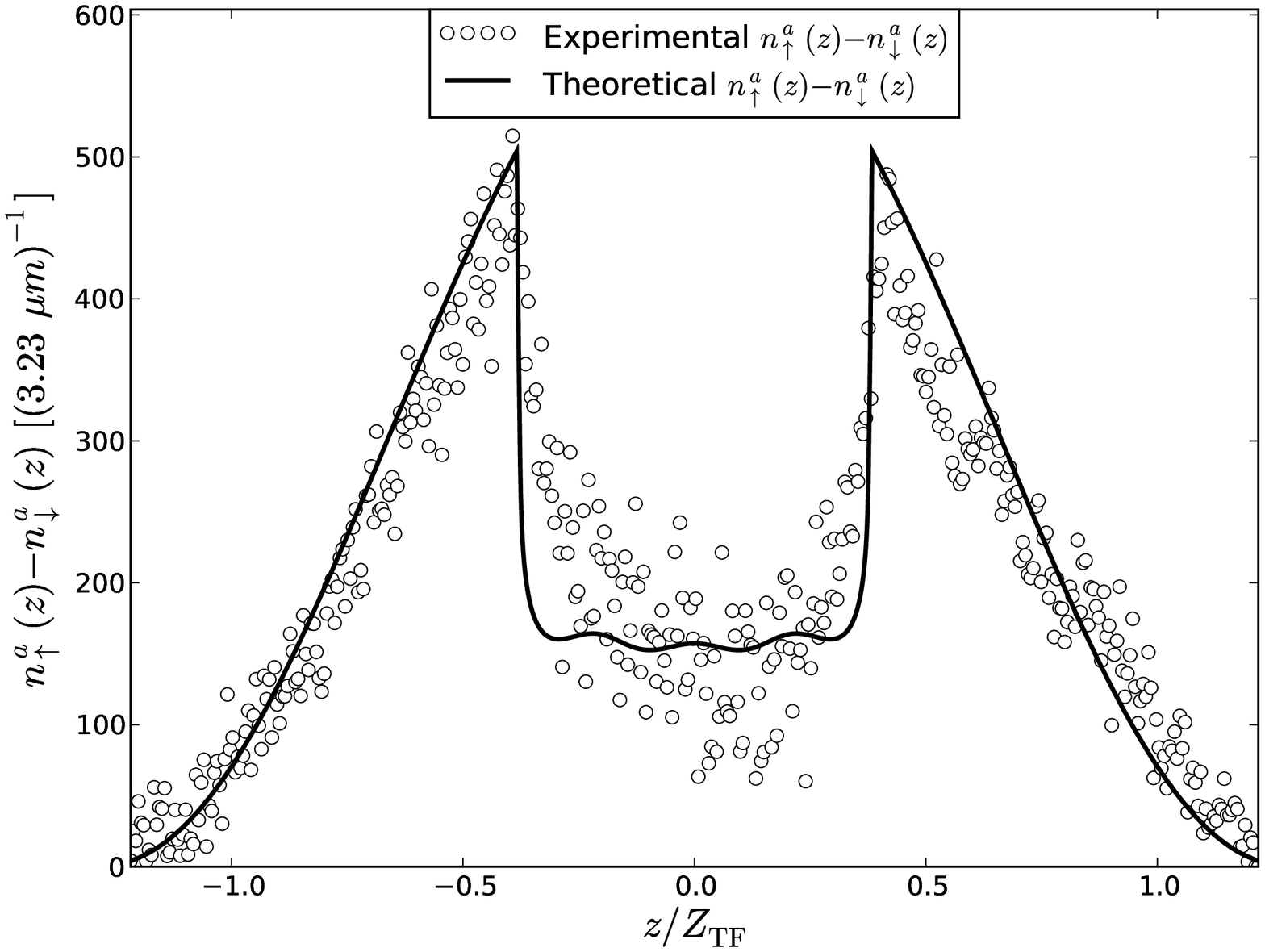}
 \hfill
 \includegraphics[width=0.48\textwidth]{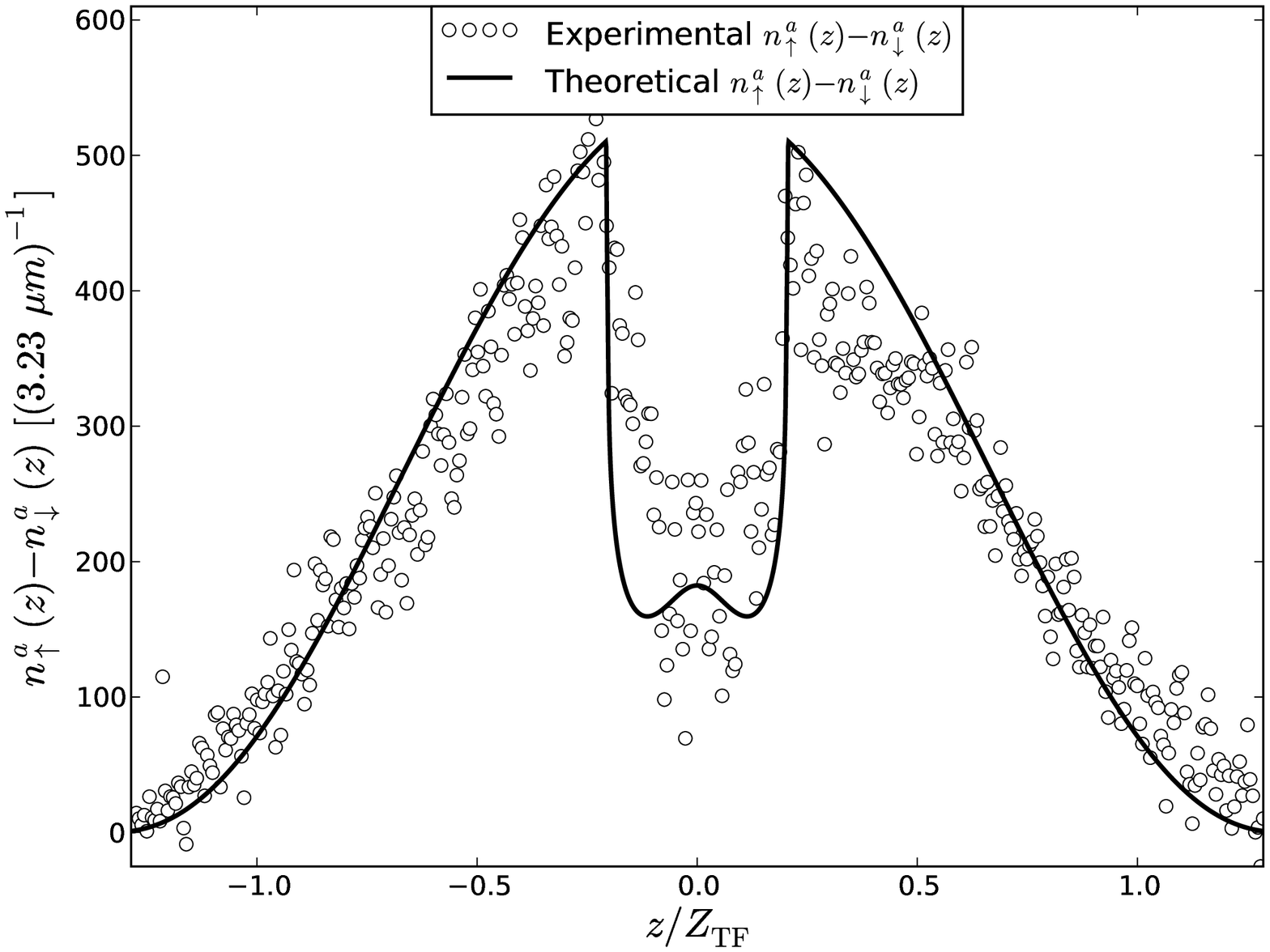}
 \caption{Axial densities.  Symbols: experimental one-dimensional $^6$Li spin densities and density differences for $P=0.39$ $(N_\uparrow=155,000,N_\downarrow=68,500)$ (left column) and $P=0.63$ $(N_\uparrow=123,600,N_\downarrow=28,000)$ (right column), from Ref.~\cite{Partridge2006a}, with permission. Lines: theoretical curves for $\eta=2.83$, taking a cigar shaped harmonic trap with small oscillation frequencies $\omega_z =(2\pi) 7.2$Hz and $\omega_r=(2\pi) 325$ Hz. Oscillations in the density difference within the superfluid region are artifacts of our ansatz~\eqref{eq:harmonicansatz}. To minimize noise, only experimental data inside an elliptical window was considered (see text). This aperture is visible in 
 figure~\ref{fig:inceta}. 
 }
 \label{fig:oneDplot}
\end{figure*}

Having explored the microscopic theory of the domain wall separating the superfluid and normal regions, we now investigate how the surface energy in this boundary affects the shape of a trapped gas.
We will assume that the zero temperature population imbalanced atomic system is phase separated into two regions: a central superfluid core surrounded by a normal shell.  We will take the normal state to be fully polarized (with $n_\downarrow=0$) and the superfluid state to be fully paired (with $n_\uparrow=n_\downarrow$).  As discussed in the introduction, this is an approximation.  Remarkably, the experiments at Rice university\cite{Partridge2006,Partridge2006a} are largely consistent with this ansatz, which was first introduced by Chevy~\cite{Chevy2006}.  As seen in figures~\ref{fig:inceta}-\ref{fig:oneDplot}, in these elongated clouds 
there is no experimental evidence of a partially polarized normal region between the fully paired superfluid and fully polarized normal regions.  The absence of this phase is significant: a partially polarized normal state is seen in experiments at MIT \cite{Zwierlein2006,Zwierlein2006a,Shin2008} and in more sophisticated theoretical calculations of the bulk phase diagram \cite{Lobo2006,chevy2,Bulgac2008}.  The unexplained behavior seen at Rice is presumably related to the small numbers of particles and the high aspect ratio of the cloud, but other considerations, such as the kinetics of spin transport, may also be important  \cite{Sensarma2007, Tezuka2008, ku2008,parishhuse}.

We will restrict our discussion to unitarity, where
 physics is  universal and the superfluid and surface energy densities between the superfluid and normal regions have simple forms. The equation of state of the central superfluid shell is given at $T=0$ by eq.~(\ref{sfeos}),
while the outer fully polarized normal shell obeys
\begin{equation}
 n_n(\mathbf{r}) = n_{\uparrow} = \frac{1}{6\pi^2}\left[\frac{2m\mu_n(\mathbf{r})}{\hbar^2}\right]^{3/2}.
\end{equation}
 The free energy densities of the bulk phases $f_{s,n}=-\int n_{s,n}d\mu$
can be written as
\begin{eqnarray}\label{be1}
f_{s,n}(\mathbf{r})=-\frac{2}{15 \pi^2}
\left(\frac{2m}{\hbar^2}\right)^{3/2}\zeta_{s,n}
\mu^{5/2}_{s,n}(\mathbf{r})
\end{eqnarray}
where $\zeta_{s}=1/(1+\beta)^{3/2}$, $\zeta_{n}=1/2$. Then
we calculate the total bulk energies $\Omega_{S,N}=\int_{s/n}d^3 \mathbf{r}
f_{s,n}[\mu(\mathbf{r}),h]$ by integrating the bulk energy densities over the
superfluid/normal regions. As previousy introduced, we model the surface energy density
$\sigma[\mu(\mathbf{r}),h]=\eta(\hbar^2/2m) n_s^{4/3}$, where $\eta$ is the dimensionless constant calculated in the previous section. We calculate the total
surface energy $E_{dw}=\int d^2 r \sigma[\mu(\mathbf{r}),h]$ by integrating
the surface energy density over the superfluid-normal boundary. 
Away from the superfluid-normal boundary, we assume that the system
is locally homogenous and the external harmonic trapping potential
$V_{trap}(\mathbf{r})=b_{\perp} \rho^2+b_z z^2= m
\omega_z^2(\lambda^2 \rho^2+z^2)/2$ is treated in the LDA by
introducing a local chemical potential
$\mu(\mathbf{r})=\mu_0-V_{trap}(\mathbf{r})$. 
Given that the experimental traps are formed by focussed laser beams, describing it via a harmonic potential should be viewed as an approximation.

\subsection{Calculation of boundary}\label{sec:theja2}

We make a completely general ansatz for the domain wall, 
only assuming rotational symmetry about the long axis of the trap.
We parameterize the boundary in terms of coordinates $f$ and $\theta$, which are related to the cylindrical coordinates $\rho$ and $z$ by
\begin{equation}
 \begin{split}
  \rho(\theta,f) &= R_{TF} f \cos\theta\\
  z(\theta,f) &= Z_{TF} f \sin\theta
 \end{split}
 \label{eq:generalized}
\end{equation}
where $R_{TF}=\sqrt{\mu_0/b_{\perp}}$, $Z_{TF}=\sqrt{\mu_0/b_{z}}$. The boundary is described by the function $f=F(\theta)$.  As shown in appendix~\ref{details}, the two-dimensional integrals for the free energy can then be simplified to one dimensional integrals, which can be performed numerically.

The optimal shape is found by minimizing the free energy functional $\Omega_T=\Omega_S+\Omega_N+E_{DW}$ on the space of functions $F(\theta)$ at fixed $N_\uparrow$ and $N_\downarrow$.  The constraints are imposed using Lagrange multipliers.

We expand $F(\theta)$ as
\begin{equation}
F(\theta)=\sum_{n=0}^{\infty} a_n \cos(2 n \theta)
\label{eq:harmonicansatz}
\end{equation}
which is compatible with the boundary conditions imposed by the symmetry of the problem, $F^\prime(0)=F^\prime(\pi/2)=0$. We truncated this series at a finite number of Fourier components $N_c$ and numerically minimized $\Omega_T$ with respect to $a_0,a_1,\ldots,a_{N_c}$. We find that we need to include more terms in this series when $\eta$ is larger, but for all values of $\eta$, the profiles become insensitive to $N_c$ for  $N_c \gtrsim 15$.

In figure~\ref{fig:inceta} we plot the boundary $F(\theta)$ that minimizes $\Omega_T$ for different values of $\eta$. The boundary becomes almost insensitive to $\eta$ for high surface tension. This behavior 
has two sources: (i) For large $\eta$ the ends become increasingly flat, so surface tension plays an increasingly insignificant role, (ii) the edges along the minor axis touch the edge of the majority cloud, at which point the superfluid-normal boundary changes to a superfluid-vacuum boundary and surface tension ceases to be important. Due to this ``saturation'' of the boundary shape with high $\eta$, 
and the difficulty of defining the boundary from noisy 2-D data, we find it convenient to follow references \cite{Silva2006a,Haque2007} and find $\eta$ by fitting our theoretical model to the 1-D axial densities, defined by $n^{(a)}_{\uparrow,\downarrow}(z)=\int dx\,dy\, n(x,y,z)$.  As illustrated in Fig.~\ref{fig:inceta}, we improve signal to noise by excluding data outside of an elliptical window \footnote{
We discarded all data outside of an ellipse which was chosen so that by eye only pixels with no particles in them were excluded.  The ellipse was chosen independently  for each spin state and each data set.  This windowing increased the signal to noise while reducing significant systematic biases due to the background.  For example by this measure the $P=0.39$ data has $N_\uparrow=155,000, N_\downarrow=68,500$, while without windowing $N_\uparrow=166,000, N_\downarrow=88,000$.}.
%
We find
 that $\eta\simeq 3$ gives an 
axial
density difference profile most closely matching the experimental density from Ref.~\cite{Partridge2006a} for $P=0.38$ and $P=0.63$.  
As seen in figure~\ref{fig:oneDplot}, the overall quality of the fit is quite good.  There are however distinct differences between the predictions of the model and the observed profiles.  These can largely be attributed to trap anharmonicities whose modelling is not reported here.

We also believe that the $\delta p$ dependance of $\eta$ may be important for capturing the exact shape of the domain wall.  Generically one would expect that this dependance would reduce $\eta$ at the ends of the boundary, increasing the curvature of the end-caps and making a smoother axial density.  This effect would also lead to an apparent polarization and number dependance of $\eta$.  Finally, we found some sensitivity to how we treat the background in each image.   For example, if we fit the axial density difference at $P=0.6$ without windowing the data, we find that $\eta=1$ provides a better fit.  This sensitivity can be attributed to structure in the background which persist throughout the image, even far from the cloud.

%


%

Since they are based upon identical models (just using different ansatz's for the boundary shape), the quality of our fits are very similar to the ones found by Haque and Stoof when investigating a large number of similar profiles \cite{Haque2007}.  Converting to our units, Haque and Stoof found $\eta=4.8\pm1.2$.  Their result is slightly higher than ours.  We attribute this difference to differences in fitting procedures (such as windowing the data) and to modeling of the trap.  Haque and Stoof used a more sophisticated Gaussian model for the trap, while we assumed it was harmonic.

\begin{figure}[ht!]
	\centering
		\includegraphics[width=\columnwidth]{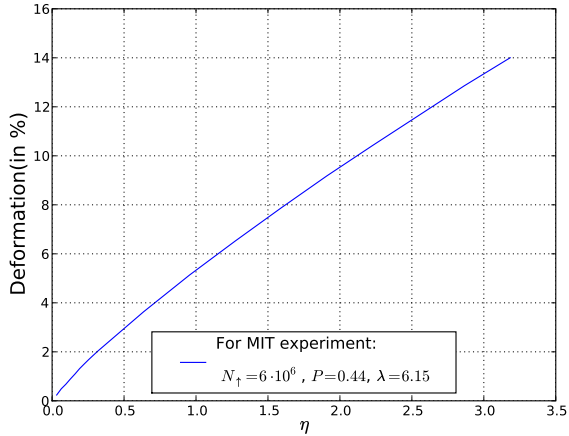}
	\caption{(Color Online) Distortion of superfluid core aspect ratio ($=1-F(\pi/2)/F(0)$) in $\%$ as a function of the dimensionless surface tension $\eta$ for parameters of Ref~\cite{Shin2008}, where $\lambda$ is the aspect ratio of the harmonic trap.}
	\label{fig:mitdeformation}
\end{figure}

The authors of Ref.~\cite{Shin2008} from the MIT experiment find no visible distortion of the superfluid cloud and quote an uncertainty of about $2\%$ for this null observation. We studied how surface tension would affect this experiment. As evident from figure~\ref{fig:mitdeformation}, a distortion of less than $2\%$ implies $\eta\lesssim1$.  Thus their null-observation is consistent with a value of surface tension on the same order as what we find in our microscopic calculation. Figure~\ref{fig:mitdeformation} shows that the surface tension needed to fit~\cite{Partridge2006a}, $\eta\simeq 3$, would cause a distortion more than 10\% in the MIT experiment (as was already pointed out in~\cite{Shin2008}), well above their threshold of detection.

\section{Summary and Conclusions}\label{sec:summary}
We presented a microscopic calculation of surface tension at a superfluid-polarized boundary in a polarized two-component Fermi gas, finding a value of  the dimensionless surface tension $\eta$ which is consistent with   an $\epsilon$-expansion~\cite{Kryjevski2008}.   We analyze both the interaction dependance and temperature dependance of $\eta$.  We argue that $\eta$ vanishes in both the BEC and BCS limits, with a peak on the BEC side of resonance near unitarity.  Furthermore, $\eta$ decreases with temperature, vanishing at the tricritical point.

We compared our microscopic calculation of $\eta$ to experimental estimates extracted by fitting a phenomenological model to axial density data.  We find that our microscopic predictions of $\eta=0.17$ at unitarity are an order of magnitude smaller than our best fits to the experimental data from  Rice~\cite{Partridge2006a} ($\eta\sim 3$).  On the other hand, phenomenological modeling of the the MIT experiment~\cite{Shin2008} bound $\eta$ to be somewhere between zero and a few times larger than our microscopic predictions.  We think that additional theoretical insight, 
as well as more experimental data, would be needed to resolve these discrepancy.  

Despite the differences in the magnitude of $\eta$, we find that the experimental observations at Rice~\cite{Partridge2006a} show all of the appropriate hallmarks of surface tension.  Using our phenomenological model, we presented a short study of how surface tension distorts the shape of the superfluid core in a harmonically trapped cloud of atoms.  We reproduce 
%
both the observed double peak structure\cite{Partridge2006} in the axial density difference profile and the distorted shape of the superfluid/normal boundary\cite{Partridge2006a}. 

We find that with increasing surface tension (figure~\ref{fig:inceta}), the superfluid/normal boundary attains a ``limiting'' shape significantly different from the isopotential contours that are predicted by the Thomas Fermi approximation. As the temperature increases, the system moves closer to the tricritical point in figure~\ref{fig:pdsketch1}, and as a result the surface tension decreases and disappears at the tricritical point in figure~\ref{fig:fig1}. As a result, surface tension-induced distortion of the superfluid core must be absent if the temperature of the
atomic trap is maintained above the tricritical point; indeed, this behavior is observed in~\cite{Partridge2006a}. We would argue that thermal effects are not responsible for the 
%
differences between experiments at Rice~\cite{Partridge2006a,Partridge2006} and MIT~\cite{Zwierlein2006,Zwierlein2006a,Shin2008}.
The strongest evidence that temperature is not the issue, is the excellent agreement
 between zero temperature Monte-Carlo results\cite{Lobo2006} and the MIT experiments~\cite{Shin2008}.

Sensarma {\it et al} \cite{Sensarma2007}, and more recently Tezuka and Ueda\cite{Tezuka2008},
attempt to understand 
 the deformation of the superfluid core by studying a microscopic model of the entire atomic cloud.  They solved the BdG equations for a relatively small number of particles (a few thousand) in  an axially symmetric system. 
 While these calculations do give more insight into the properties of these systems, it is difficult to extract quantitative information from them.  In particular, the small particle numbers lead to a much larger surface area to volume ratio than in experiments, and artificially amplifies the role of surface tension.  
 We believe that those calculations do not ``explain" the inconsistencies which we observe between the magnitude of surface tension in our microscopic BdG  calculations and our phenomenological modeling of the experimental data.  It is possible that larger simulations of that sort might be more useful in this regard.

\section{Acknowledgments}
We thank Stefan Natu, Kaden Hazzard, and Mohit Randeria for useful discussions.   We thank Wenhui Li and Randall Hulet for discussions, critical comments and  providing us the raw experimental data from~\cite{Partridge2006a}. This work was partially supported by NSF grant numbers PHY-0456261, PHY-0758104, and by the Alfred P. Sloan Foundation.

\appendix

\section{Cutoff dependance of the BdG calculations}\label{stefans_appendix}
In order to compute the surface tension across the BEC-BCS crossover we used a numerical calculation based on the BdG equations~\eqref{BdGequations}. Here we give details about our numerical approach, showing that we have used sufficiently large cutoffs to produce unbiased results.

We find parameters such that bulk normal and superfluid have the same free energy and then minimize the functional~\eqref{mean_field_free_energy} with respect to the order parameter for a configuration with a domain wall between the two phases. The excess free energy of this configuration is attributed to the domain walls, and allows us to extract a surface tension. 
%

We assume that the order parameter varies only along the $z$-direction. In the simulation we discretize this one dimensional space on a uniform grid with $N$ grid points and approximate the gradients through a fourth order finite difference matrix. We find it convenient to use periodic boundary conditions in the $z$-direction, simulating two domain walls.  We have verified that the interaction between the two walls is negligable.

Translational invariance in the directions perpendicular to the interface(i.e. $x$- and $y$-directions) implies that the $u$ and $v$'s are of the form
\begin{eqnarray}
u(\mathbf{r})=e^{i \mathbf{k}_{\perp} \mathbf{r}_{\perp}} u_n(z) \hspace{1cm} v(\mathbf{r})=e^{i \mathbf{k}_{\perp} \mathbf{r}_{\perp}} v_n(z)
\end{eqnarray}
One has to solve a $2N\times 2N$ matrix eigenvalue problem for each $k_\perp=|\mathbf{k}_\perp|$. The coupling constant $U$ depends on the UV-cutoff and was fixed by renormalizing $\Omega_{MF}(\Delta_0)$ through the result from a direct calculation~\eqref{homogenous_mean_field_free_energy} where a uniform order parameter was assumed. We found that when seeded with an order parameter profile with two domain walls 
the minimization algorithm converges to 
a local minimum with two domain walls.  This solution correctly obeys the self-consistent gap equation.

\begin{figure}[ht!]
	\centering
	\includegraphics[width=\columnwidth,angle=0]{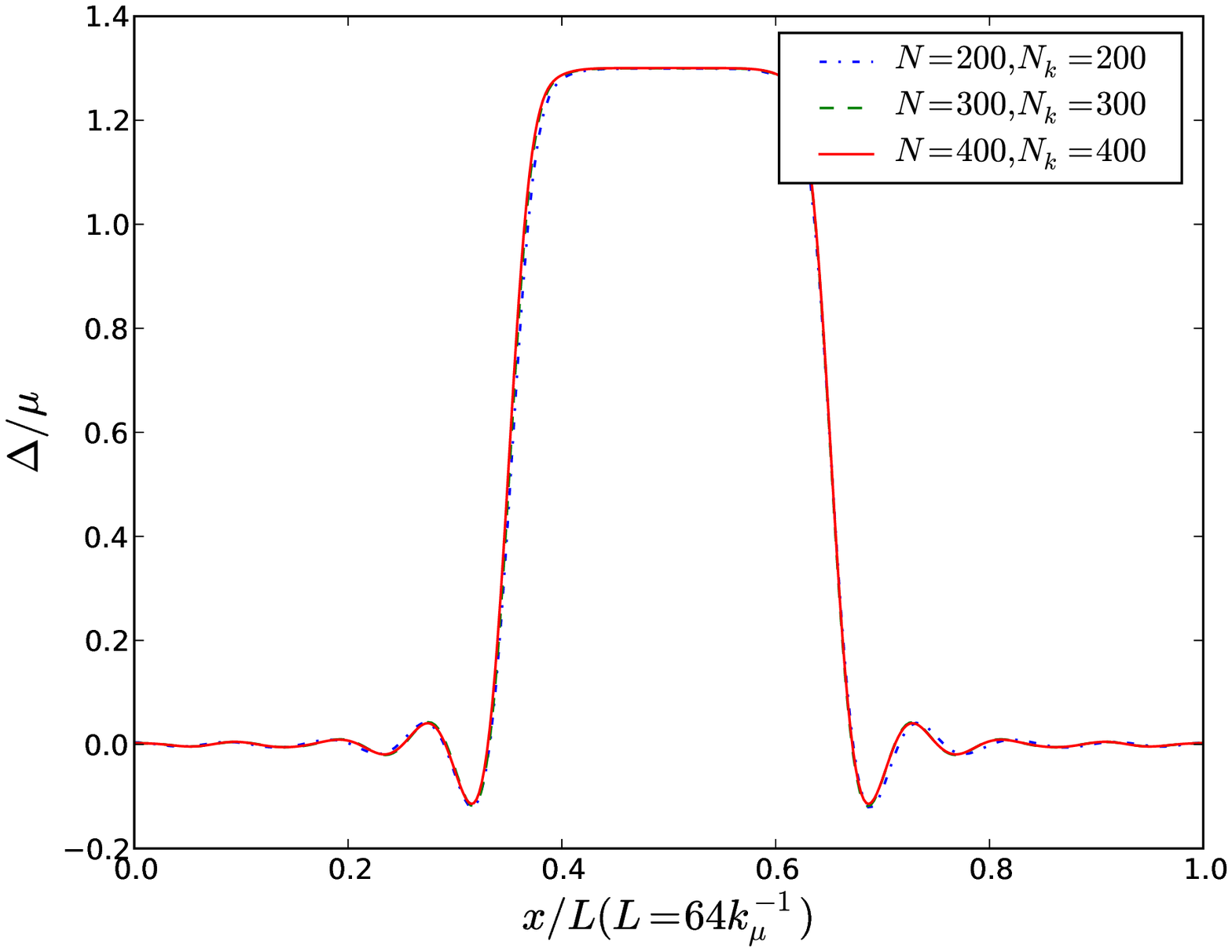}
	\includegraphics[width=\columnwidth]{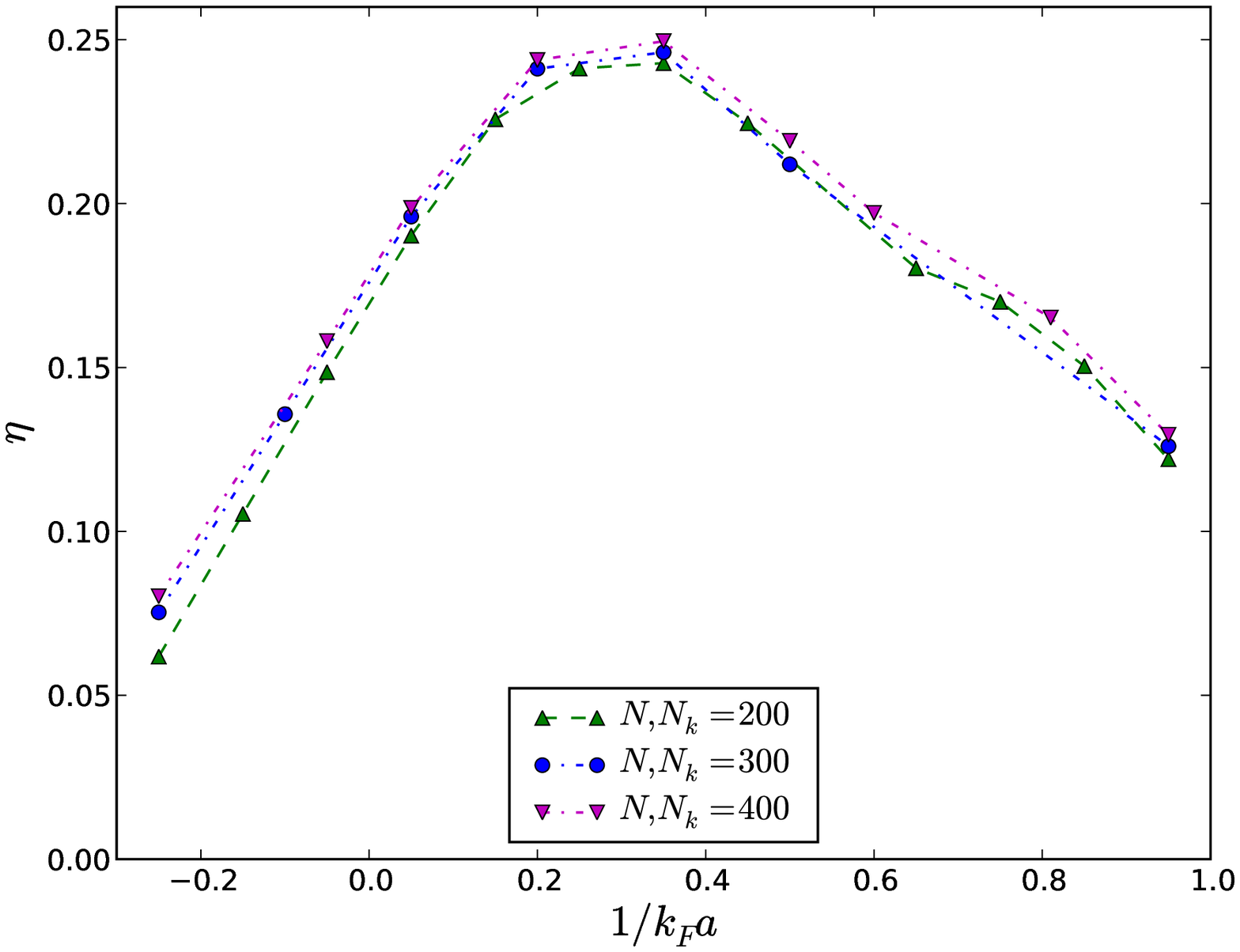}
	\caption{(Color Online) {\bf Top:} Representative order parameter profiles for different cutoffs computed using the BdG equations at $1/k_F a=0.05$. For better visibility a line connecting the data points is displayed. {\bf Bottom:} Dimensionless surface tension constant $\eta$ for different cutoffs as a function of $1/k_F a$. (Color online)}
	\label{fig:stscale}
\end{figure}

We systematically increased $N$ to check the convergence of the order parameter profiles and surface tension (see Fig.~\ref{fig:stscale}). In the deep BEC limit, the gradient expansion  becomes a good approximation to the BdG equations, enabling us to check our BdG calculation.  In that limit we found excellent agreement between these theories, giving us confidence in the accuracy of our numerical methods.
The largest number of grid points $N$ and transverse modes $N_k$ we used were $N=400$ and $N_k=400$; these were, for example, used to obtain Fig~\ref{fig:st}. We feel that any residual errors from our finite gridpoints/cutoffs are much smaller than the errors introduced through the mean field approximation.

\section{Details of gradient expansion}\label{sourish_appendix}

This appendix contains technical details of the gradient expansion used for finding the domain wall energy at finite temperature.

We calculate the uniform gap ($\Delta_0$) by solving
\begin{equation}
\int_0^\infty dy\,\left[1-\frac{\sinh E_y}{\cosh E_y+\cosh\beta h}\frac{y^2}{E_y}\right]=0
\end{equation}
where $E_y=\sqrt{(y^2-\beta\mu)^2+(\beta\Delta)^2}$. We numerically perform the integral up to a cutoff $y_c$, then analytically approximate the remainder by expanding the integrand to sixth order in $1/y$.
We calculate the free energy of the solution from (\ref{mean_field_free_energy}), again breaking the integral into two pieces, and analytically integrating the tail. We then find the parameters for which the normal and superfluid states have equal energy.

We write $(2 k_B T/V) \chi(q)={\cal  A} \tilde\chi(q \sqrt{\beta/2m})$, with ${\cal A}=1/(2\pi^2)(2m/\beta)^{3/2}$ to express
\begin{eqnarray}
\tilde\chi(q) &=&\int_0^\infty dk\,\frac{k\mathcal{L}(k,q)-2k^2-\frac{1}{2}q^2+2\beta\mu}{k^2-\beta\mu+\frac{1}{4}q^2}\\
\mathcal{L}(k,q) &=& \frac{1}{q}\log(Z_-^\uparrow Z_-^\downarrow/Z_+^\uparrow Z_+^\downarrow)-2 k\\
Z_{s}^\sigma&=&
1+e^{(k-s q/2)^2-\beta\mu_\sigma}.
\end{eqnarray}
After tabulating this integral, calculating the energy of a domain wall is straightforward within the error function ansatz of Eq.~ (\ref{eq:erf_ansatz}).  Again, we split all integrals into two parts, integrating the tails analytically.

To extract $\eta$ we also calculate the bulk superfluid density,
\begin{equation}
\begin{split}
n_s =& \frac{1}{2\pi^2}\left(\frac{2m}{\hbar^2\beta}\right)^{3/2}\int_0^\infty dy\, y^2\times\\
&\left[1-\left(\frac{y^2-\beta\mu}{E_y}\right)\frac{\sinh E_y}{\cosh E_y + \cosh\beta h}\right],
\end{split}
\end{equation}
using $\Delta=\Delta_0$.

\section{Evaluation of phenomenological Free energy}\label{details}
This appendix gives the analytic expressions used to calculate the
free energy of an arbitrary domain wall.  We parameterize the boundary by $f=F(\theta)$, in terms of which the coordinates of the boundary are $ \rho(\theta,f) = R_{TF} f \cos\theta$ and
$ z(\theta,f) = Z_{TF} f \sin\theta$.  The surface energy is
\begin{equation}
\begin{split}
E_{DW} &= A_{dw}\int \left(1-\frac{\rho^2}{R^2_{TF}}-\frac{z^2}{Z^2_{TF}}\right)^2 d^2 r\\
&= 2A_{dw}\int_0^{\pi/2}d\theta F(\theta)\cos\theta \\
&\times [F^\prime(\theta)\cos\theta-F(\theta)\sin\theta]\times [1-F(\theta)^2]^2
\\
&\times
\sqrt{1+\biggr(\frac{Z_{TF}}{R_{TF}}\biggr)^2\biggr(\frac{F^\prime(\theta)\sin\theta+F(\theta)\cos\theta}{F^\prime(\theta)\cos\theta-F(\theta)\sin\theta}\biggr)^2}
\end{split}
\end{equation}
where we define the coefficient
\begin{equation}
A_{dw} = \hbar\omega_z R_{TF}^2Z_{TF} \left[\frac{2m}{\hbar^2}\right]^{3/2}\left[\frac{\eta\pi\mu_0^{3/2}}{(1+\beta)^2 (3\pi^2)^{4/3}}\right].
\label{eq:boundary_coeff}
\end{equation}
We write the free energy of the superfluid core as
\begin{eqnarray}
\Omega_{S} &=& A_{s}\int_s \rho d\rho dz
\left(1-\frac{\rho^2}{R^2_{TF}}-\frac{z^2}{Z^2_{TF}}\right)^{5/2}\nonumber\\
&=& 2A_{s}\int_0^{\pi/2} d\theta \cos\theta\int_0^{F(\theta)}df f^2(1-f^2)^{5/2}\nonumber\\
&=&  2A_{s}\int_0^{\pi/2}d\theta G_1[F(\theta)]\cos\theta\\
 A_{s} &=& -\zeta_s R_{TF}^2Z_{TF} \left[\frac{2m}{\hbar^2}\right]^{3/2} \left[\frac{4 \mu_0^{5/2}}{15\pi}\right]
 \label{eq:sf_coeff}\\
  G_1(x) &=& \frac{1}{384}\Bigl[x\sqrt{1-x^2}\left(-15+118x^2-136x^4+48x^6\right)\Bigr.\nonumber\\
&& +\Bigl.15\sin^{-1}(x)\Bigr]
\end{eqnarray}
Similarly, the free energy of the fully polarized normal shell, $\Omega_{N} = A_{n}\int_n \rho d\rho dz (1+\gamma-\rho^2/R_{TF}^2-z^2/Z_{TF}^2)^{5/2}$, is:
\begin{eqnarray}
\Omega_{N}&=&
2A_{n}(1+\gamma)^4\biggr[\frac{5\pi}{256}-\int_0^{\pi/2}d\theta
G_1\biggr[\frac{F(\theta)}{\sqrt{1+\gamma}}\biggr]\cos\theta\biggr]\nonumber\\
 A_n &=& -\zeta_n R_{TF}^2Z_{TF} \left[\frac{2m}{\hbar^2}\right]^{3/2} \left[\frac{4 \mu_0^{5/2}}{15\pi}\right].
 \label{eq:eq38}
\end{eqnarray}
The total number of atoms in the two phases are given by
\begin{equation}
\begin{split}
\label{eq:atomnumber}
N_{s} &= 2B_{s}\int_0^{\pi/2} d\theta \cos\theta G_2[F(\theta)]\\
N_{n} &= 2B_n(1+\gamma)^3\left[\pi/32- \int_0^{\pi/2}d\theta \cos\theta G_2\left[\frac{F(\theta)}{\sqrt{1+\gamma}}\right]\right]
\end{split}
\end{equation}
where 
\begin{equation}
 \begin{split}
  B_{s,n} &= \zeta_{s,n} \frac{2}{3\pi}\left[\frac{2m\mu_0}{\hbar^2}\right]^{3/2} R_{TF}^2 Z_{TF}\\
  G_2(x) &= \frac{1}{48}\left[x\sqrt{1-x^2}(-3+14x^2-8x^4)+3 \sin^{-1}(x)\right].
 \end{split}
\end{equation}
Thus both the free energy, and the constraints of fixed $N$ and $P$ reduce to one dimensional integrals.


\end{document}